\documentclass{article}
\usepackage[english]{babel}
\usepackage[utf8]{inputenc}
\usepackage{johd}
\usepackage{ragged2e}
\usepackage{amsmath}
\usepackage{array}
\usepackage{graphicx}
\usepackage{caption}
\usepackage{booktabs}
\usepackage{amssymb}
\usepackage{subfig}
\usepackage{natbib}
\usepackage[export]{adjustbox}
\usepackage{wrapfig}
\title{Beyond the Traditional VIX: A Novel Approach to Identifying Uncertainty Shocks in Financial Markets}

\author{Ayush Jha$^{1}$$^{*}$, Abootaleb Shirvani$^{2}$, Svetlozar T. Rachev$^{3}$ and Frank J. Fabozzi$^{4}$ \\
        \small $^{1}$Department of Economics, Texas Tech University \\
        \small $^{2}$Department of Mathematical Sciences, Kean University \\
        \small $^{3}$Department of Mathematics and Statistics, Texas Tech University \\
        \small $^{4}$Carey Business School, Johns Hopkins University \\\\
        \small $^{*}$Corresponding author: Ayush Jha; \tt{ayush.jha@ttu.edu} \\
}
\date{}

\begin{document}
\maketitle
\begin{abstract} 
\noindent We introduce a new identification strategy for uncertainty shocks to explain macroeconomic volatility in financial markets. The Chicago Board Options Exchange Volatility Index (VIX) measures market expectations of future volatility, but traditional methods based on second-moment shocks and time-varying volatility of the VIX often fail to capture the non-Gaussian, heavy-tailed nature of asset returns. To address this, we construct a revised VIX by fitting a double-subordinated Normal Inverse Gaussian L\'{e}vy process to S\&P 500 option prices, providing a more comprehensive measure of volatility that reflects the extreme movements and heavy tails observed in financial data. Using an axiomatic approach, we introduce a general family of risk-reward ratios, computed with our revised VIX and fitted over a fractional time series to more accurately identify uncertainty shocks in financial markets. \end{abstract}

\noindent\\


\section{Introduction}
\justifying
\setlength{\parskip}{10pt}

\citet{bloom} introduced the concept of an uncertainty shock, examining stock market volatility and its impact on broader economic outcomes. In his work, he identifies second-moment shocks—uncertainty shocks—as a primary channel through which stock market volatility affects the economy, observed through both realized and implied option prices. Bloom uses the CBOE Volatility Index (VIX) to quantify this uncertainty, which reflects market expectations of future volatility over the next 30 days based on S\&P 500 options prices. However, analysis of S\&P 500 returns reveals that these option prices follow a non-normal distribution. As a result, measuring market volatility using only the second moment, or standard deviation, is insufficient, as it assumes a symmetric distribution and fails to account for the “heavy tails” that are known to be present in market data (\citet{Cont}). This limitation weakens the standard deviation's ability to capture the full impact of extreme market fluctuations, leading to underestimating the true effect of such shocks. 

To effectively capture uncertainty, the identification strategy must consider the impact of extreme events on option prices, as these events amplify uncertainty through what is known as tail risk. \citet{KOV} highlights the importance of tail risks in cases where data are non-normally distributed. By introducing "disaster risk" as a factor, they show how uncertainty is magnified through heavy tails. Unlike normal distributions, which have thin tails and suggest a low probability of extreme events, non-normal distributions with heavy tails better represent the likelihood of rare, impactful events. Additionally, the symmetry of a normal distribution implies equal chances of extreme positive or negative outcomes—“disasters” and “miracles”—a pattern not reflected in observed data.

In the context of financial time series data, \citet{Cont} points out that the non-Gaussian nature of the distribution of asset returns (first moments) makes a strong case for using other measures of dispersion to observe the variance of the returns. The large movements in financial markets, identified via the heavy tails of the distribution, cannot be regarded as simple outliers in the sample. Therefore, amplification by heavy tails motivates us to model the tails of the distribution of asset returns appropriately. In addition, modeling the heavy tails and skewness of the distribution will enable us to examine the magnitude of the uncertainty caused by large movements in the financial market. 

Given the emphasis on heavy tails in non-Gaussian stable distributions, using second moments to capture the volatility as a measure of uncertainty will not factor in the extreme tail risks, making the notion of uncertainty insufficient. Moreover, volatility does not account for the directional bias of the uncertainty, rendering uncertainty shocks non-explanatory of large movements in the financial market. Risk--reward ratios (R/R) offer a more nuanced approach, considering that they are widely used as performance measures in financial decision-making. R/R ratios focus on the tail risk to explain the amplification in uncertainty shocks by including the rewards (potential gains on the right tail) and risks (potential losses on the left tail). 

Furthermore, since risk--reward measures can help reduce max drawdowns, they offer practical insights into managing financial downturns, which are integral to understanding uncertainty shocks amplified by the heavy tails of the financial time series distribution. In addition, to capture the observed volatility in financial markets, we use a double subordinated Normal Inverse Gaussian (NIG) L\'{e}vy process to construct a revised VIX from S\&P 500 option prices. This NIG L\'{e}vy process explains the skew and fat-tailed properties of index option prices and gives rise to an arbitrage-free, option pricing model which is then used to compute the in-sample, implied volatility of the stock market. 

By combining the newly constructed revised VIX with computed R/R ratios, we produce a new series of uncertainty shocks, which better explain the variations caused by large market movements. These uncertainty shocks are robust in explaining the unexplained amplification of uncertainty through tail risks by combining the direction of the potential gains and losses of the stock market given the heavy tails and skewness of the option (asset) prices. The revised measure of volatility, which identifies uncertainty shocks, has broad economic implications for analyzing market risks and responses. To the point of all information embedded, given the new measure captures intrinsic time volatility, to correctly capture the skewness and fat tails of the S\&P500 index, it can explain the impact of monetary policy tightening (e.g., federal funds rate hikes or forward guidance) on risky assets such as the index itself. It will capture the effects surrounding pre- and post-announcements of policies made by the Federal Reserve.

In addition, it can explain the shift in risk-neutral density implying fat tails due to geopolitical risk in the form of armed conflicts (e.g., the Russia-Ukraine, or the Iran-Israel conflict). The transmission channel will reflect the impact of such events on the agent's risk aversion and willingness to resort to the third derivative of their utility function to trigger precautionary savings. Similarly, other rare disasters' impact on volatility in the financial markets, like COVID-19, will also be explained by the revised VIX given it captures all information about market activity in the event of rare disasters. Moreover, it has implications for sector-specific or firm-level friction in the form of the present value of future cash flow being heavily discounted if firm earnings are lower than market expectations (for instance, tech stocks being sold at a large mass after tech firms failed to meet market expectations in early 2022 on account of their earnings). 

This research question aims to contribute to numerous strands of the literature on macro-volatility in financial markets and the channels of uncertainty. \citet{bloomfu} argued that investors want to be compensated for higher risk, and because greater uncertainty leads to increasing risk premia, this should raise the cost of finance. Hence, capturing an increase in risk premia can be attributed to the time-intrinsic and fat tails characteristics of the uncertainty explaining the magnitude of extreme events. \citet{kelly-jiang} pointed out that researchers have hypothesized that heavy-tailed shocks to economic fundamentals help explain certain asset pricing behavior that has proved otherwise difficult to reconcile with traditional macrofinance theory.

\citet{rietz} was among the first to emphasize the phenomenon of fat tails attributed to the rare disaster hypothesis. Moreover, \citet{bansal-yaron} construct a model of long-run risks that incorporates fat-tailed endowment shocks. \citet{duffie-pan-singleton} explain the mechanics of modeling extreme events with jump-diffusions to capture heavy-tailed events. Extending on this, \citet{Shirvani-Stoyanov-Fabozzi} extend \citet{mehra-prescott}'s framework to account for the growth rate of consumption and dividends that follow a fat-tailed distribution. The authors assume that the log-growth rates of consumption and dividends follow a NIG distribution and find that the constant relative risk aversion (CRRA) estimate derived from the NIG-fitted model is significantly lower than the estimate obtained using a log-normal distribution fit. 

Furthermore, the tail risk mechanism that accentuates uncertainty channels is highlighted in \citet{orlik-veldkamp} explaining the fluctuations in the macroeconomic measure of uncertainty. In addition, \citet{kvv} identifies that agents are unaware of the true distribution of shocks but use data to estimate it non-parametrically. However, transitory, especially extreme events, generate persistent changes in beliefs and macro outcomes. While these studies explain the extreme events channel of uncertainty, intrinsic time volatility to explain heavy tails remains an ambiguous territory. This paper is the first to explain the subordination process using the intrinsic time-volatility and the heavy tails of the asset return to capture the dynamics of the financial market and using R/R ratios to estimate uncertainty shocks as macro-volatility in financial markets.

Section 2 presents the double subordinated NIG L\'{e}vy process for S\&P 500 option prices used to construct the revised VIX. Section 3 presents the double subordinated NIG L\'{e}vy process European call option pricing used to price the S\&P500 returns following a Normal Double Inverse Gaussian (NDIG) log-price process. Section 4 presents the method used for estimating the parameters. Section 5 demonstrates the robustness of multiple subordinated models of volatility. Section 6 presents the general families of risk--reward ratios. Section 7 presents the novel identification strategy for uncertainty shocks. Section 8 concludes the paper.

\section{Double Subordinated NIG L\'{e}vy process for S\&P 500 option prices}
\setlength{\parskip}{10pt}
\justifying

The double subordinator framework in our approach to construct a revised VIX involves a L\'{e}vy subordinator process. We define the functional form of the stochastic process following a L\'{e}vy process as described by \citet{carr} and \citet{bitcoin-shirvani-rachev}. In dynamic asset pricing theory, the price dynamics explain the behavior of the risky financial asset. Consider the price process $S_{t}, t \in [0,\tau]$, where $t<\infty$ is the time horizon, implying that $\tau$ is the maturity date of a financial contract. Therefore, a L\'{e}vy process $\mathbb{T} = (T_{t}, t \geq 0, T_{0} = 0)$ with non-decreasing trajectories or sample-paths is known as a L\'{e}vy subordinator. To begin with, we start by modeling S\&P500 options in a standard Black--Scholes--Merton option pricing model assuming normality. The price process of the risky asset is

\begin{equation}
    S_{t}^{(BSM)} = e^{X_{t}^{(BSM)}},\;\;\;\;t \in [0,\tau]
\end{equation}
\vspace{-0.5cm}
\begin{equation}
     X_{t}^{(BSM)} = X_{0} + \mu_{1t} + {\sigma_{1}}B_{t},\;\;\; \mu_{1} \in R, \sigma_{1} > 0, X_{0} = ln(S_{0}), S_{0} > 0 
\end{equation}

\noindent where $X_{t}$ is the log-price process and $\mathbb{B} = (B_{t},\;t \geq 0)$ is a standard Brownian motion. Moreover, knowing that the option prices of S\&P 500 are non-normally distributed (in particular, having heavy tails), we can follow \citet{mandelbolt-taylor} and \citet{clark} who suggested the use of a subordinated Brownian motion, where the price process $S_{t}^{(ss)}$ and thereby, the log price process is defined by

\begin{equation}
    S_{t}^{(ss)} = e^{X_{t}^{(ss)}},\;\;\; t \in [0,\tau]
\end{equation}
\vspace{-0.5cm}
\begin{equation}
    X_{t}^{(ss)} = X_{0} + \mu_{2,t} + \sigma_{2}B_{T_{t}}\:\:,\;\;\; \mu_{2} \in R, \sigma_{2} > 0
\end{equation}
where $\mathbb{T} = (T_{t}, t \geq 0, T_{0} = 0)$ is a L\'{e}vy subordinator.

\citet{Shirvani-Rachev-Fabozzi} describes the properties of various multiple subordinated log-return processes designed to model leptokurtic asset returns, showing that multiple subordinated log-asset return processes can imply heavier tails than single subordinated models, and thus have the ability to capture the third moment (skewness) and the fourth moment (kurtosis). Hence, a double subordination process to model the fat tails of S\&P 500 option prices may be an appropriate choice. 

Let $S_{t}$ denote the price process of the S\&P 500 options, where

\begin{equation}
    S_{t} = e^{X_{t}},\:\:t\in[0,\tau]
\end{equation}
\vspace{-0.5cm}
\begin{equation}
    X_{t} = X_{0} + \mu_{3t} + {\gamma}U_{t} + {\rho}T(U_{t}) + \sigma_{3}B_{T(U_{t})}\:,\;\;t \geq 0, \mu_{3} \in R, \sigma_{3} > 0, X_{0} = ln(S_{0}), S_{0} > 0 
\end{equation}
and the members of the triplet (${B_{s}, T_{s}, U_{s}},\;\; s \geq 0$) are independent processes generating a stochastic basis $(\Omega, \mathcal{F}, \mathbb{F} = (\mathcal{F}_{t}, t \geq 0), \mathbb{P})$ denoting continuous time preferences. We refer to {$B_{s},\;\; s \geq 0$} as a standard Brownian motion along with {$T_{s}, s \geq 0, T_{0} = 0$} and {$U_{s}, s \geq 0, U_{0} = 0$} being the L\'{e}vy subordinators. Moreover, $B_{t}$, $T_{t}$, and $U_{t}$ are $\mathcal{F}_{t}$ adapted processes whose trajectories are right-continuous with left limits. \citet{Shirvani-Rachev-Fabozzi} denote the double subordinated process by $T(U(t)),\;t\geq0$. Therefore, Eq. (4) is a double-subordinated log-price process.

 Considering the S\&P500 options prices are modeled as a subordinated geometric Brownian motion, a multiple subordinated model would sustain the bivariate semi-martingale structure. In addition, L\'{e}vy subordinators (volatility intrinsic time and heavy tails) do not disrupt the S\&P500's time-changed process and sustain the bivariate semi-martingale property under a risk-neutral measure consistent with the fundamental asset pricing theorem. Consider the case where $T(t)$ and $U(t)$ are inverse Gaussian L\'{e}vy processes, i.e., $T(1) \sim IG(\lambda_{T}, \mu_{T})$ having the pdf

 \begin{equation}
     f_{T(1)}(x) = \sqrt{\frac{\lambda_{T}}{2{\pi}x^3}}exp{\frac{\lambda_{T}(x-\mu_{T})^2}{2\mu_{T}^2x}},\;\;\;\;x\geq0,\;\mu_{T}>0,\;\lambda_{T}>0
 \end{equation}

Consider the second subordinator, $U(1) \sim  IG(\lambda_{U}, \mu_{U})$. Given the nature of both subordinators, we can regard $X_{t}$ in Eq. (4) as the \textit{Normal Double Inverse Gaussian} (NDIG) log-price process. Therefore, the characteristic function (c.f.) of $X_{1}$ is defined by

\begin{equation*}
    {\psi}x_{1}(v) = \mathbb{E}[e^{iv{x_{1}}}]
\end{equation*}

Therefore, $\mathbb{E}[e^{iv{x_{1}}}]$ can be expanded and written as:

\begin{equation}
\mathbb{E}\left[e^{iv{x_{1}}}\right] = e^{\left(iv\mu_{3} + \frac{\lambda_{U}}{\mu_{U}}\left[1 - \sqrt{1 - \frac{2\mu_{U}^2}{\lambda_{U}}\left(\frac{\lambda_{T}}{\mu_{T}}\left(1 - \sqrt{1 - \frac{\mu_{T}^2}{\lambda_{T}}\left(2iv\rho - \sigma_{3}^2v^2\right)}\right)\right)} + iv\lambda\right]\right)}
\end{equation} 

given $v \in \mathbb{R}$. Hence, the moment generating function (MGF) of $X_{1}$ is $M_{X_{1}}(w) = \mathbb{E}[e^{wX_{1}}]$, where $w \in \mathbb{R}$. Therefore, the NDIG process modeled in Eq. (6) and the expansion of the c.f. modeled in Eq. (8) has eight parameters, i.e., $\mu_{3}, \sigma_{3}, \gamma, \rho, \mu_{\tau}, \mu_{T}, \lambda_{\tau}, \lambda_{T}$. Given the complexity of fitting the processes modeled to the data, we will follow the methodology explained by \citet{bitcoin-shirvani-rachev} where six parameters can be estimated from the model. Still, the remaining two are computed by taking expectations. 

\begin{figure}[H]
    
    \begin{minipage}[b]{0.49\textwidth}
        \centering
        \includegraphics[width=\textwidth]{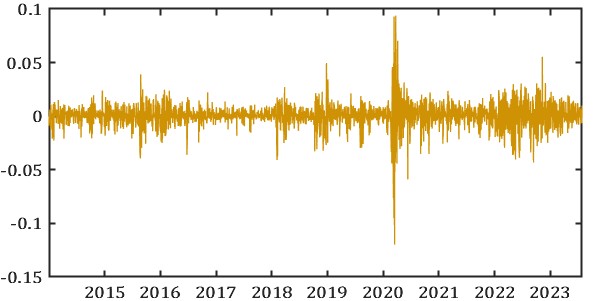}
        \caption{S\&P500 Options Prices}
    \end{minipage}
    \hfill
    \begin{minipage}[b]{0.49\textwidth}
        \centering
        \includegraphics[width=\textwidth]{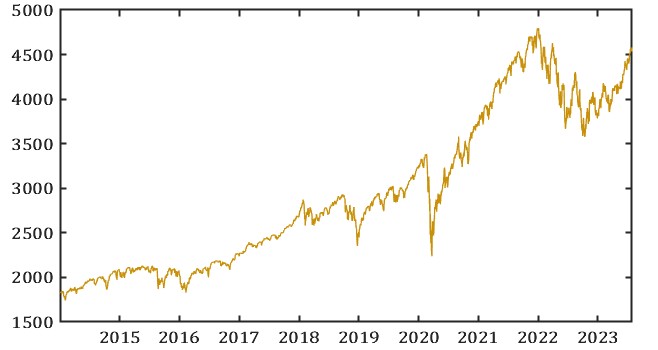}
        \caption{Last Price (S\&P500)}
    \end{minipage}
    \label{1}
\end{figure}


\section{NDIG Model for European Call Option Pricing}
\setlength{\parskip}{10pt}
\justifying

We assume complete markets where agents can continuously trade a risky asset and a risk-free bond. The NDIG model will be used to price a European Call Option, $\mathcal{C}$, with the underlying risky asset, $\mathcal{S}$, being the options prices of the S\&P500. The discounted price process $e^{-r_{t}}S_{t}$ with $r \geq 0$ being the risk-free rate needs to sustain the martingale structure, therefore, we need to derive an Equivalent Martingale Measure (EMM) $\mathbb{Q}$ of $\mathbb{P}$ on the stochastic base $(\Omega, \mathcal{F}, \mathbb{F} = (\mathcal{F}_{t}, t \geq 0), \mathbb{P})$. We follow the same methods as described in \citet{bitcoin-shirvani-rachev} of using a Minimal Conditional Martingale Measure (MCMM), hence, pricing European options under this measure satisfies the no-arbitrage condition.

\begin{figure}[H]
    
    \begin{minipage}[b]{0.49\textwidth}
        \centering
        \includegraphics[width=\textwidth]{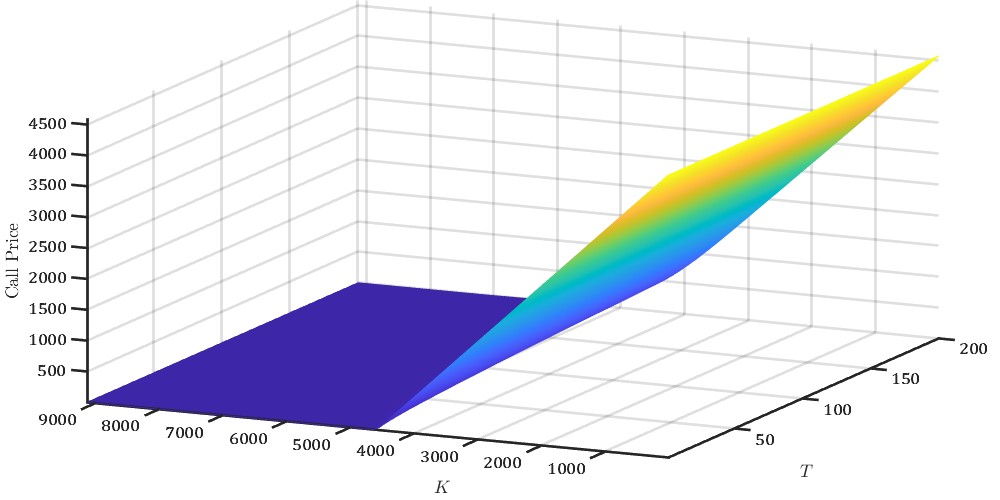}
       
    \end{minipage}
    \hfill
    \begin{minipage}[b]{0.49\textwidth}
        \centering
        \includegraphics[width=\textwidth]{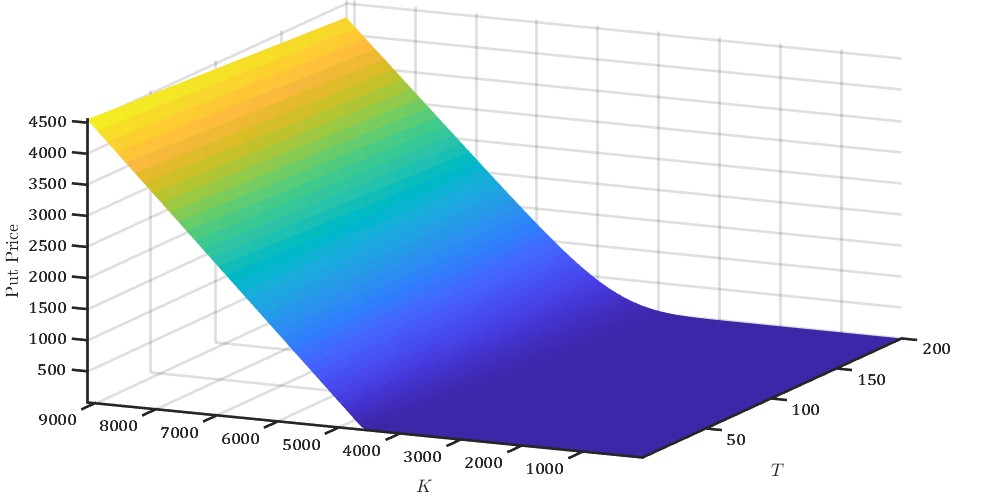}
        
    \end{minipage}
    \label{2}
    \caption{Implied Volatility Surface. (i) Call Price and (ii) Put Price}
\end{figure}

In addition, given that the c.f. of $X_{t}$ is defined in Eq. (8), we follow the \citet{carr-madan} approach of using the Fast Fourier Transform (FFT) to price options in cases where the c.f. of the log-price of the underlying asset is known analytically. Consequently, the inital point for applying the FFT is derived via the following relation:

\begin{equation}
    C(S_{0},\;r,\;k,\;\tau) = \frac{e^{-rr-ak}}{\pi}\int_{0}^{\infty} e^{-ivk}\frac{{\psi}_{lnS_{T}^{(\mathbb{Q})}}(v - i(a + 1))}{a^2 + a - v^2 + i(2a + 2)v} \,dv
\end{equation}

\citet{carr-madan} show that the numerical solutions yield an `optimum value' for $`a'$ and control over the error produced by truncating the integral in Eq. (9) over a finite domain $[0,\;v_{max}]$. Furthermore, to pin down $a_{max}$, the upper bound on $a$, we can follow the process outlined in Eq. (10) and compute the upper bounds in moving windows to investigate the behavior of $a$ over time. Figure (4) analyzes the behavior of $a_{max}$. For instance, it is equipped to describe the large movement in the financial market in March 2020, indicated by a significant drop in the value of the upper bound. 

\begin{equation}
    a_{max} = \frac{1}{\sigma_{3}^2}\sqrt{\rho^2 + \lambda_{U}\left(1 - \frac{\lambda_{U}}{4\lambda_{T}}\sigma_{3}^2\right)} - \frac{\rho}{\sigma_{3}^2} - 1
\end{equation}

Moreover, option prices for $\mathcal{C}$ are determined using FFT over Eq. (9) with multiple values of the strike price $K$ and maturity time $T$. To maintain stability in the implied volatility surface of the S\&P500 options, we place a restrictive condition on $a$, namely, $a < 1$.

\begin{figure}[H]
    \centering
    \includegraphics[width=0.7\linewidth]{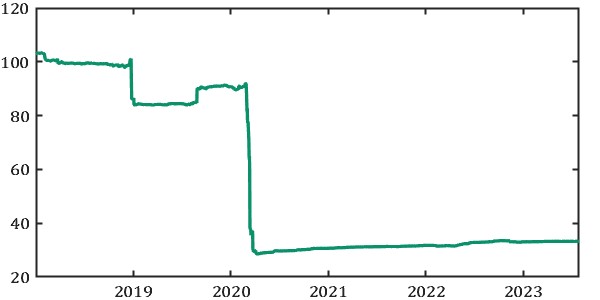}
    \caption{Values for the upper bound $a_{max}$ computed in the time period starting 01/02/2014 to 07/28/2023.}
    \label{3}
\end{figure}

Figures (5) and (6) compare the kernel density fits of the S\&P500 options. As hypothesized, the density fit of S\&P500 options showcases the fat tails that cannot be captured through a normal distribution. Comparing the NDIG fit to the Student's $t$-distribution, we can see that NDIG is better at capturing the heavy tails in the data previously not accounted for when measuring uncertainty. Therefore, the method of using two subordinators to explain the volatility through the intrinsic time of the asset return process and the heavy tails of the data is an appropriate strategy.
\vspace{0.5cm}
\begin{figure}[H]
    
    \begin{minipage}[b]{0.49\textwidth}
        \centering
        \includegraphics[width=\textwidth]{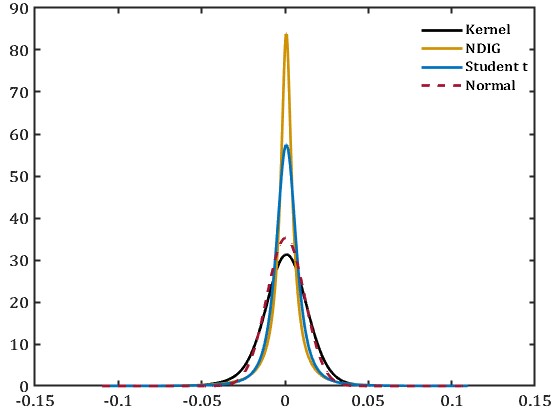}
        \caption{Density Fit}
    \end{minipage}
    \hfill
    \begin{minipage}[b]{0.48\textwidth}
        \centering
        \includegraphics[width=\textwidth]{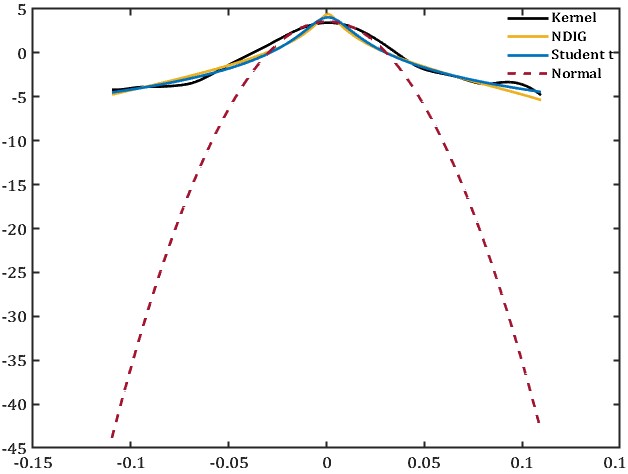}
        \caption{Log Scaled Density Fit}
    \end{minipage}
    \label{4}
\end{figure}

\section{Parameter Estimation}

To estimate the parameters of the double subordinated log-price process, we fit the NDIG model described in the previous two sections to the S\&P500 daily returns ranging from January 2, 2014 up to July 28, 2023. The NDIG model is explained in Eqs. (5), (6), (7), and (8), and is used to estimate six parameters: $\theta = (\mu_{3},\;\sigma_{3},\;\mu_{U},\;\lambda_{U},\;\mu_{T},\;\lambda_{T})$. We work with the assumption that the subordinators $T(t)$ and $U(t)$ are used to model the skewness and heavy tails (kurtosis) of the asset return process and the intrinsic time of the asset return time series, respectively. 

Therefore, the estimation of the first four moments follows the same method as described in \cite{bitcoin-shirvani-rachev} to estimate parameters using the NDIG model for Bitcoin log returns. 

\begin{equation}
    \min_{\mu_{3},\;\sigma_{3},\;\rho,\;\lambda_{T},\;\lambda_{U}} ({\Delta}M_{1})^2 + ({\Delta}M_{2})^2 + ({\Delta}M_{3})^2 + ({\Delta}M_{4})^2 + ({\Delta}CF)^2
\end{equation}

The given minimization problem in Eq.~(11) is subject to the following five constraints to calculate the five choice variables or parameters using first-order conditions. The constraint for the first moment is $\sim$ $({\Delta}M_{1})^2 = 1 - \frac{\mathbb{E}[X]}{\mathbb{E}[p_{t}]}$. The constraint for the second moment is $\sim$ $({\Delta}M_{2})^2 = 1 - \frac{Var[X]}{Var[p_{t}]}$. The constraint for the third moment is $\sim$ $({\Delta}M_{3})^2 = 1 - \frac{Skew[X]}{Skew[p_{t}]}$. The constraint for the fourth moment is $\sim$ $({\Delta}M_{3})^2 = 1 - \frac{Kurt[X]}{Kurt[p_{t}]}$. The constraint for the c.f. is $\sim$ $({\Delta}CF)^2 = \int_{-\infty}^{\infty} \left(\frac{1}{n} \Sigma_{j=1}^{n}e^{ivx_{j}} - {\psi}x_{t}(v,\;\theta)\right)^2 dv$. Here, $p_{t}$ denotes the asset return time series observed from the S\&P500 options prices. Moreover, $({\Delta}CF)$ depends on the one-to-one correspondence between the cdf and the c.f. as noted in \citet{bitcoin-shirvani-rachev}. We follow the method described in \citet{yu-chf} to estimate the integral of the c.f. We get the following parameter estimates using the method of moments and empirically fitting the c.f. These estimates are presented in Table 1.
\vspace{0.3cm}

\begin{figure}[h]
    \centering
    \includegraphics[width=0.7\linewidth]{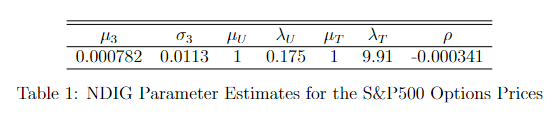}

\end{figure}

In addition, our objective of interest leads us to estimate the parameters in the four moving windows to examine the behavior of the first four moments of the data. While $\mu_{U}$ and $\mu_{T}$ remain on a constant path throughout the period used in the estimation of the parameters, we see significant movements in the drift term $\mu_{3}$, the volatility parameter $\sigma_{3}$, $\rho$, $\lambda_{U}$ and $\lambda_{T}$. Figure (7) shows the rolling window parameter estimates. 

\begin{figure}[H]
    
    \begin{minipage}[b]{0.47\textwidth}
        \centering
        \includegraphics[width=\textwidth]{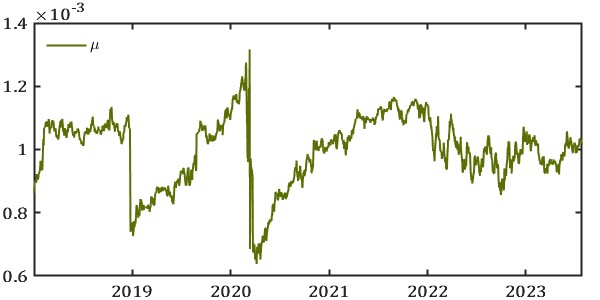}
        
    \end{minipage}
    \hfill
    \begin{minipage}[b]{0.47\textwidth}
        \centering
        \includegraphics[width=\textwidth]{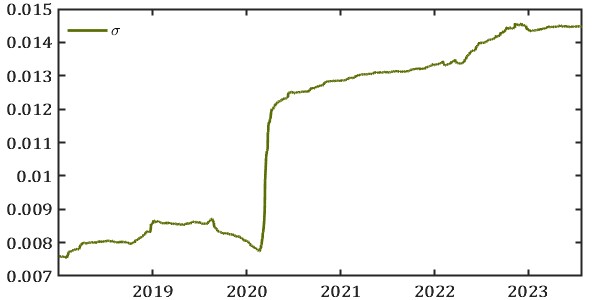}
        
    \end{minipage}
    \hfill
    \begin{minipage}[h]{0.47\textwidth}
        \centering
        \includegraphics[width=\textwidth]{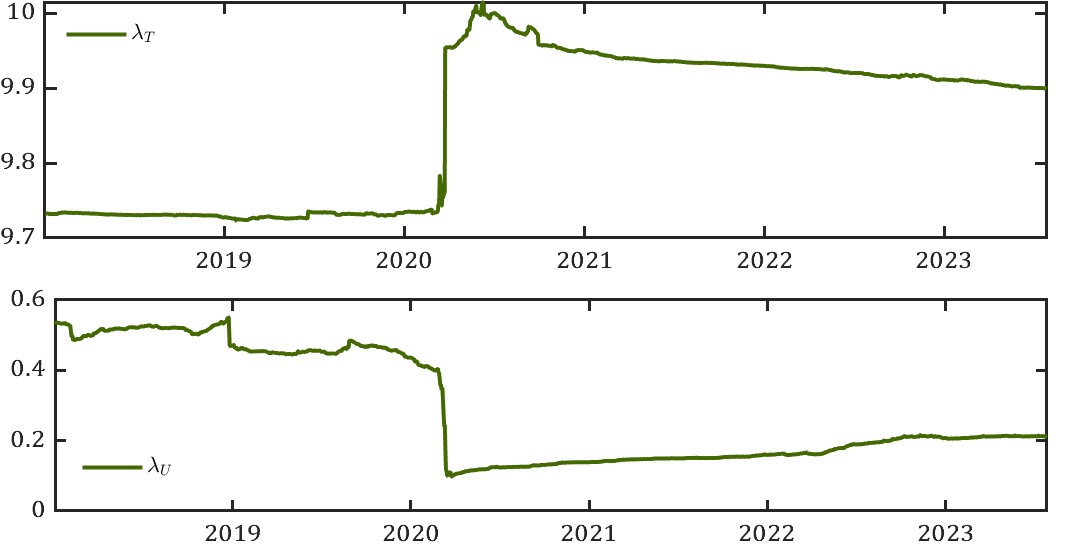}
       
    \end{minipage}
    \hfill
    \begin{minipage}[h]{0.47\textwidth}
        \centering
        \includegraphics[width=\textwidth]{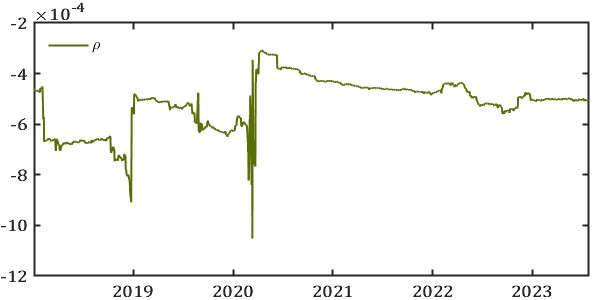}
        
    \end{minipage}
    \label{5}
    \caption{Parameter estimates in the 4-year moving window: (i) $\mu_{3}$, (ii) $\sigma_{3}$, (iii) $\lambda_{T}\;\&\;\lambda_{U}$, and (iv) $\rho$.}
\end{figure}

\section{Multiple Subordinated Models of Volatility}
\setlength{\parskip}{10pt}
\justifying

The VIX, a product of the CBOE, is often called the ``fear gauge'' of the market and measures the market's expectation of volatility over the next 30 days implied by S\&P500 option prices. Most of the literature on uncertainty shocks focuses on capturing volatility in the financial market using the VIX standard deviation or incorporating time-varying volatility clustering as shocks into macroeconomic models. But, first, using the standard deviation works only when the underlying data assumes a symmetric distribution: yet it is evident from Section 2 that the VIX follows a non-normal distribution. This leads to an underestimation of the true magnitude of uncertainty shocks from the VIX. Second, a popular approach to computing uncertainty shocks is examining the time-varying volatility clustering of the VIX. 

Time-varying or stochastic volatility models capture the variability of the VIX over time. Since uncertainty is not constant, models such as the Generalized Autoregressive Conditional Heteroskedasticity (GARCH) and its variant, the Fractionally Integrated GARCH (FIGARCH) are used to capture the time-varying nature of volatility clustering. These models account for the tendency of high-volatility events to be followed by more high-volatility, while low-volatility periods are often followed by low volatility. By analyzing this clustering in the VIX, it is possible to identify periods with increased uncertainty. 

However, one of the key identifiable issues in computing the volatility surfaces of the VIX is using local volatility models. These models assume a deterministic function of current asset price and time $t$ and use it to compute the value of the volatility surfaces based on near-term and next-term expirations of the S\&P500 options\footnote{Refer to eq (12) which describes the expiration times and methodology of computing the value of the VIX.}. Given that local volatility models are deterministic functions and not stochastic functions which account for past observed asset mean returns and volatility, they fail to reconcile with the fundamental theorem of asset pricing, thereby, making it inconsistent with the dynamics of the financial market. Moreover, there are two reasons why local volatility measures are inferior to multiple subordinated models, which in our case, accounts for stochastic volatility and heavy tails of the data distribution.

Consider an agent who can continuously trade a risky asset and risk-free bond in their portfolio assuming that markets are complete and asset prices clear all markets. There are no financial frictions in the market and agents observe the information about the assets available at time $t$. The first reason why local volatility models fail is that having two sources of uncertainty, namely the time-varying nature of the asset return process and the fat-tails and skewness of options prices, the agent trading a risky asset and a riskless bond cannot hedge their risk perfectly (\citet{ssrf}), thus succumbing to one channel of risk out of the two. 

\begin{figure}[h]
    
    \begin{minipage}[b]{0.49\textwidth}
        \centering
        \includegraphics[width=\textwidth]{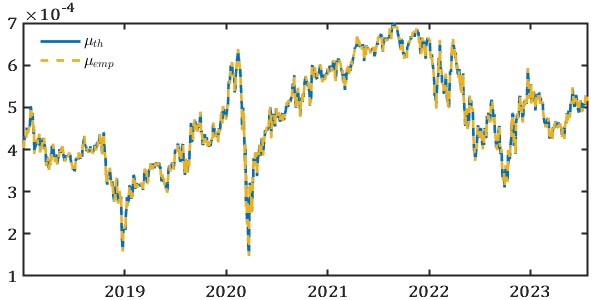}
        
    \end{minipage}
    \hfill
    \begin{minipage}[b]{0.49\textwidth}
        \centering
        \includegraphics[width=\textwidth]{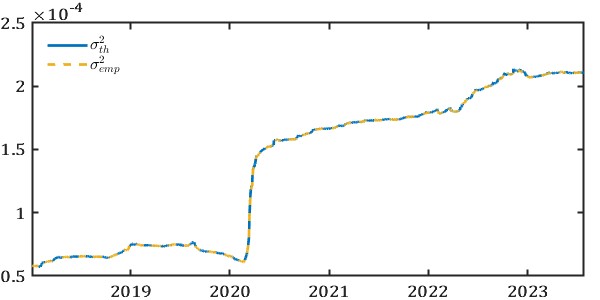}
       
    \end{minipage}
    \hfill
    \begin{minipage}[h]{0.49\textwidth}
        \centering
        \includegraphics[width=\textwidth]{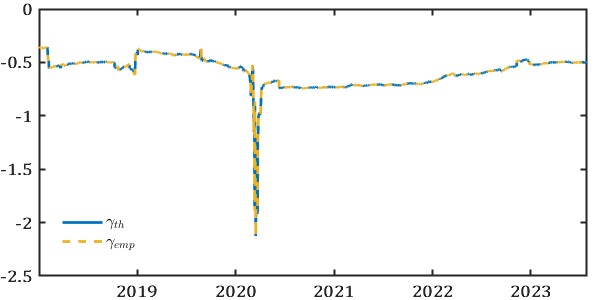}
        
    \end{minipage}
    \hfill
    \begin{minipage}[h]{0.49\textwidth}
        \centering
        \includegraphics[width=\textwidth]{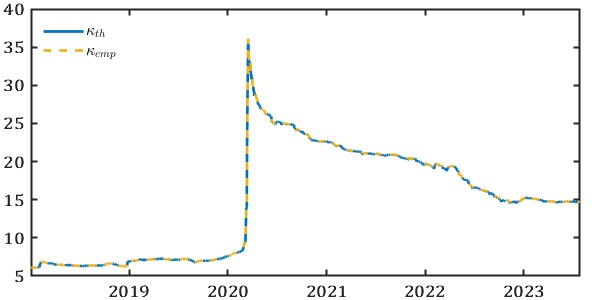}
        
    \end{minipage}
    \label{6}
    \caption{Comparison of the four moments: Theoretical vs. Empirical $\sim$ (i) $\mathbb{E}[X]$, (ii) $Var[X]$, (iii) $Skew[X]$, and (iv) $Kurt[X]$. `Th': theoretical moments estimated from fitted parameters. `Emp': empirical moments computed from the S\&P500 option prices.}
\end{figure}

The second reason why local volatility models fail is that they are not semi-martingales\footnote{A semi-martingale is a type of stochastic process that plays a central role, particularly in the modeling of asset prices. Semi-martingales are general enough to include many important classes of processes (such as Brownian motion and L\'{e}vy processes) while still allowing the use of stochastic calculus. See \citet{Meyer-Dellacherie}} implying that the fundamental theorem of asset pricing fails. \citet{delbaen-sch} exposit the fundamental theorem of asset pricing for unbounded stochastic processes. The theorem states that the absence of arbitrage possibilities for a stochastic process $\mathcal{S}$ is equivalent to the existence of an equivalent martingale measure for $\mathcal{S}$. Let $\mathcal{S} = (S_{t})_{t \in R_{+}}$ be an $\mathbb{R}^d$-valued semi-martingale defined on the stochastic basis $(\Omega, \mathcal{F}, (\mathcal{F}_{t})_{t \in R_{+}}, \mathbb{P})$. Then $\mathcal{S}$ satisfies the condition of `No Free Lunch with Vanishing Risk'\footnote{This condition essentially states that in a well-functioning financial market, it is impossible to construct a trading strategy that yields a risk-free profit with zero initial investment and no risk of loss. This concept is closely related to the absence of arbitrage opportunities in the market. See \citet{delbaen-sch-2}} if and only if there exists a probability measure $\mathbb{Q} \sim \mathbb{P}$ such that $\mathcal{S}$ is a sigma-martingale with respect to $\mathbb{Q}$. 

As a result, the local volatility model does not generally satisfy the semi-martingale property due to its deterministic nature which does not reconcile with the fundamental theorem of asset pricing. Instead, the price process may exhibit characteristics such as explosions or other paths, and therefore, a local volatility model would prevent a decomposition into a local martingale plus a finite variation process. One way to address the issue of this failure of local volatility measures is to assume that a proxy for volatility (like VIX) is a tradable asset. This way, the market is complete, and the pricing model becomes a bivariate semi-martingale (\citet{ssrf}). Therefore, multiple subordinated processes can be applied to a bivariate semi-martingale (for instance, a stochastic process including bivariate cases), and the resulting process will generally retain the semi-martingale property (\citet{Barndorff-Nielsen-Shephard}), depending on the characteristics of the subordinators.

Furthermore, \citet{Shirvani-Rachev-Fabozzi} introduced the intrinsic time volatility or volatility subordinator model to reflect the heavy-tail phenomena present in asset returns. They studied the question of whether the VIX is a volatility index that adequately reflects the intrinsic time and showed that the volatility index fails to appropriately capture the intrinsic time for the SPDR S\&P 500. The VIX, as a measure of time change, does not reflect all the information required to correctly capture the skewness and the fat tails of the S\&P 500 index. Hence, an NDIG L\'{e}vy process model with a time-varying volatility subordinator adequately accounts for the measure of intrinsic time.


\citet{kelly-jiang} developed a tail risk measure that is correlated with the tail risk measure extracted from S\&P500 options and negatively predicts real economic activity. In their methodology, they correctly explain that the dynamic tail risk estimates are infeasible in a univariate time series model because of the infrequent nature of extreme events. However, one of the major drawbacks of this paper is the authors fail to explain how to handle the family of extreme negative thresholds, $u_{t}$. For instance, if we estimate the tail parameter and find that the tail is excessively heavy and wish to purchase insurance on the portfolio consisting of assets with returns $R_{i}$, the method falls short. They also fail to distinguish whether the returns considered are logarithmic or arithmetic, which is crucial for tail estimation. In addition, they do not discuss the idea of portfolio insurance (option pricing) when the tails have a Pareto distribution. For instance, the agent cannot estimate the risk using their benchmark model and proceed to buy puts as portfolio insurance instruments using the Black--Scholes--Merton model since it assumes Gaussian-ness (thin tails), leading to risk estimations away from the true estimates.

\citet{orlik-veldkamp} and \citet{KOV} provide extensions to tail risk estimations to compute uncertainty shocks. However, they only explain one channel of uncertainty, that of the heavy tails embedded in the idea of the rare disaster hypothesis. Consequently, the idea of a semi-martingale with heavy-tailed behavior raises additional concerns. While symmetric Pareto distributions are infinitely divisible, they do not support option pricing because the Esscher transform\footnote{The Esscher transform is used to price risky assets and derivatives. See \citet{esscher}.} requires an exponential moment. Our approach solves this problem and provides a more robust framework for risk assessment and option pricing. The approach of using NDIG L\'{e}vy processes allows us to estimate the risk of a portfolio (or individual assets) consistently with the Fundamental Asset Pricing Theorem (\citet{duffie}). 

Based on the estimation power of the multiple subordinated models to account for volatility measures, we estimate the historical volatility of the VIX and report it along with the second moment from the NDIG estimates. Figure (9) presents both results in a 1008 rolling window based on the sample period of the data used for the estimation. To calculate the value of the VIX index from S\&P500 option prices, we use the same index formula that the CBOE specifies:

\begin{equation}
    VIX = 100\;*\;\sqrt{W_{1}\sigma_{1}^2 + W_{2}\sigma_{2}^2}
\end{equation}

In Eq. (12), 1 and 2 refer to near-term and next-term option expiration times. While near-term means option contracts expiring in 23--30 days, next-term means contracts expiring in 31--37. The weights specified by $W_{j}$, where $j = 1,\;2$, denote the expiration times post normalization accurate to the minute. We impose the following constraints on the weights: $0 \leq W_{1}$ and $W_{2} \leq 1$ along with $W_{1} + W_{2} = 1$. \citet{bitcoin-shirvani-rachev} provide the functional form of the weights (which we follow) along with an expression describing the near-term and next-term volatilities captured by $\sigma_{j}$.

\begin{figure}[H]
    \includegraphics[width=1\linewidth]{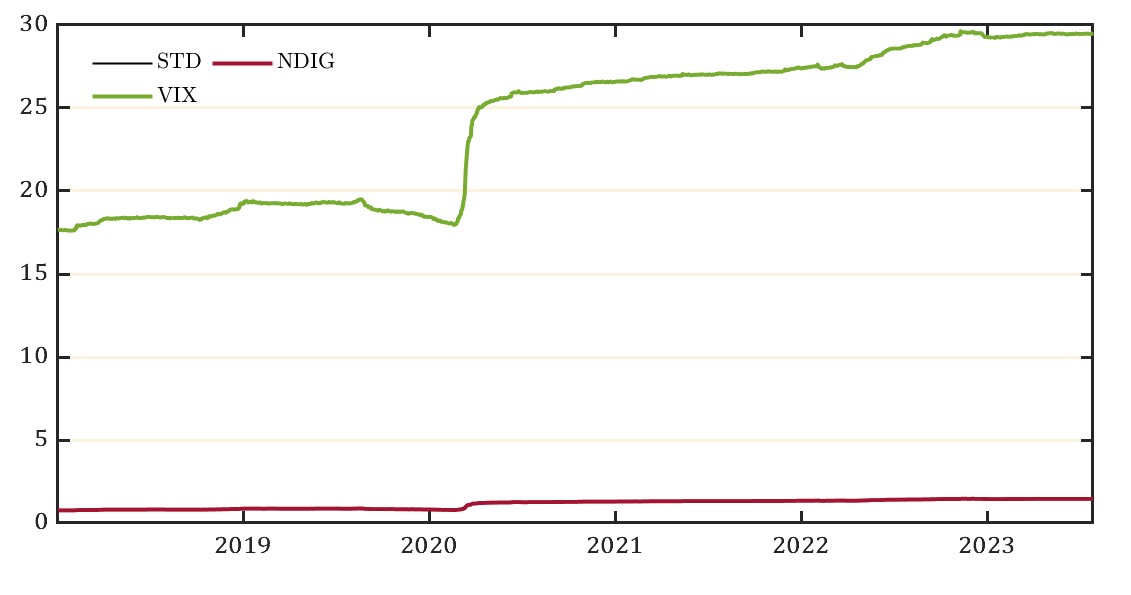}
    \caption{Historical Volatility of VIX \& NDIG Estimates in 1008 Rolling Windows}
    \label{7}
\end{figure}

To compute the values of the revised VIX based on Eq. (12) where the option prices are modeled using the NDIG log price process, we use the multiple subordinated model to calculate the prices of the European call and put options with near- and next-term expirations. Therefore, by using the subordinated method, we derive a revised measure of volatility using the parameters estimated by the method of moments. The volatility of a unit increase of $X_{t}$, the log-price process of S\&P500, is the NDIG volatility that accounts for both the geometric Brownian motion and the two L\'{e}vy subordinators of the model. 

Furthermore, the rolling NDIG parameters are estimated from the S\&P500 daily returns from January 2, 2014, up to July 28, 2023. We specify the window size to be 1008 by accounting for 252 trading days in a year and choosing four years as our moving window. Hence, we derive the annualized volatility implied by the NDIG volatility. To match the scale of the estimates of historical volatility and the intrinsic-time volatility implied by the NDIG model, we linearly scale the estimates given in Figure (9) by reducing the mean to zero and scaling the variance to 1. Figure (10) presents the normalized Volatility of VIX as our revised VIX index that explains the volatility of the financial market using intrinsic time volatility and heavy tails of the distribution.  

Figure (10) measures the historical volatility and matches it with the intrinsic time volatility, and it can be seen there that the revised index accurately captures the jumps and diffusions in the markets previously unaccounted for and therefore crucial in estimating the uncertainty as a macro-volatility in financial markets. There are two key facts to note about the revised VIX index. First, we see a jump in March 2020 which captures the large crash in the S\&P500 daily returns post the heightened uncertainty about economic conditions following the impact of the news of a global pandemic. Considering the impact generated by persistent volatility, the NDIG model preserves the volatility measure implied by the intrinsic time subordinator.

\begin{figure}[H]
    
    \includegraphics[width=1\linewidth]{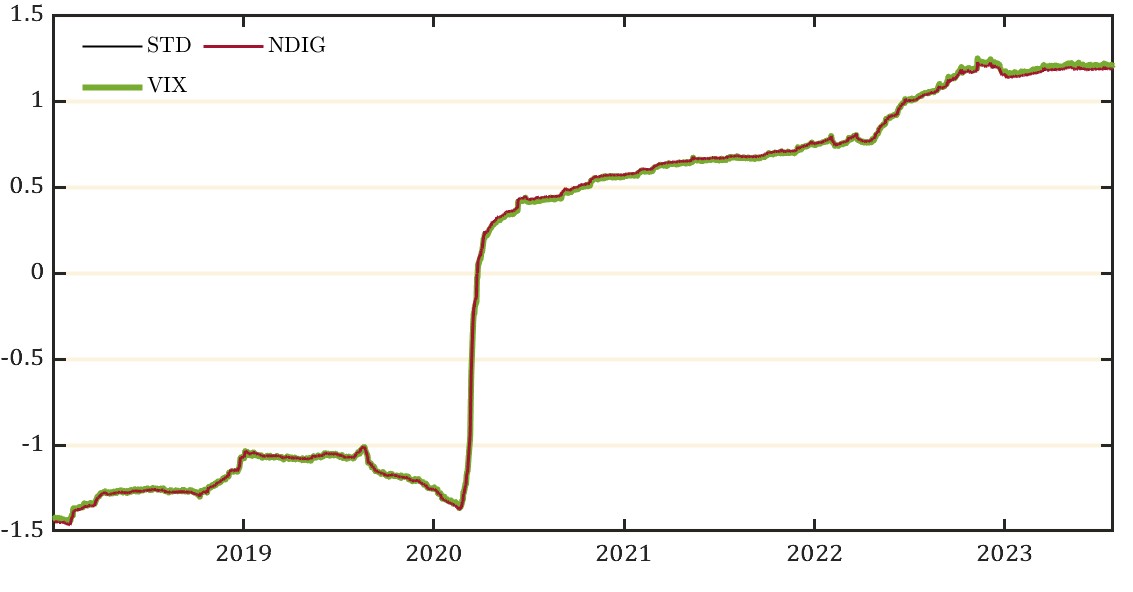}
    \caption{Normalized Volatility to Match NDIG Estimates: \textbf{Revised VIX} \{\textbf{Volatility of VIX (VVIX)}\}}
    \label{8}
\end{figure}

 Second, following the path of persistent volatility post-pandemic, we see another jump in the volatility following the events that characterized the plummets in early 2022 following consistent hikes in the federal funds rate, the fear of the start and continued geopolitical conflict between Russia and the Ukraine, along with the tech stock selloff due to an unexpected fall in tech firms' earnings indicated by the reports of their earnings.

\section{Risk--Reward Ratios over Fractional Time Series}

R/R ratios offer a balanced approach to exploring the potential gains and losses in the financial market due to violent market movements. These measures help address the asymmetry in risk perceptions and the potential for large losses, and are thereby helpful in extracting meaningful signals from the volatility noise that are not accounted for when using measures of dispersion over symmetric distributions. Using an axiomatic approach, every performance measure or R/R ratio should satisfy the properties of, first monotonicity, which means that more is better than less. Second, quasi-concavity leads to preferences that value averages higher than extremes, encouraging diversification. Third, scale invariance and last, being distribution-based. 

Let $\mathcal{X}$ be a convex set of random variables on a probability space $(\Omega, \mathcal{F}, \mathbb{P})$. Each element $X \in \mathcal{X}$ denotes a financial return over time length $T \in \mathbb{R}_{+}$. Given these conditions, consider an R/R ratio of the following form:

\begin{equation}
    \alpha(X) = \frac{\theta(X)^+}{\rho(X)^+}
\end{equation}

\noindent for a reward measure $\theta: X \rightarrow \mathbb{R} \cup {\{\pm\infty\}}$ and a risk measure $\rho: X \rightarrow \mathbb{R} \cup {\{\pm\infty\}}$. In addition, $x^+$ denotes $max{\{x,0\}}$ and $x^-$ denotes $- min{\{x,0}\}$. The ratio $\alpha(X)$ should satisfy the following two conditions:

\begin{enumerate}
    \item \textbf{(M)} Monotonicity: $\alpha(X)\geq\alpha(Y)\:\forall\:X,Y\in \mathcal{X}$ such that $X{\geq}Y$
    \item \textbf{(Q)} Quasi-Concavity: $\alpha({\lambda}X + (1-\lambda)Y) > min({\alpha}(X), {\alpha}(Y)) \; \forall\; X,Y \in \mathcal{X}\; and \;\lambda \in \mathbb{R}$ such that $0 \leq \lambda \leq 1$. 
\end{enumerate}

\citet{cheridito-kromer} explain that monotonicity is a minimal requirement that every performance indicator should satisfy. It simply implies that more of a financial return is better than less and preferred by all agents. Moreover, quasi-concavity has can explain the aversion to uncertainty. If $\alpha$ is monotonic and quasi-concave, averages are preferred to extremes and diversification is encouraged. In cases when $\alpha$ does not satisfy the required properties, there are $X, Y \in \mathcal{X}$ and a scalar $\lambda \in (0, 1)$ such that $\alpha({\lambda}X + (1-\lambda)Y) < min({\alpha}(X), {\alpha}(Y))$. In such a case, research on Value-at-Risk (VaR) \citet{artzner} shows that there will be a concentration of risk. 

Moreover, there is a large family of R/R ratios that also satisfy the following conditional properties:
\vspace{-0.5cm}
\begin{enumerate}
    \item \textbf{(S) Scale-Invariance:} $\alpha({\lambda}X) = \alpha(X)\;\forall X \in \mathcal{X}\;and \;\lambda\in\;\mathbb{R}_{+} \backslash
    \{0\}\;such\;that\;\\{\lambda}X \in \mathcal{X}$
    \item \textbf{(D) Distribution-based:} $\alpha(X)$ only depends on the distribution of $X$ under $\mathbb{P}$.
\end{enumerate}

Given that performance ratios should satisfy the first two mandatory properties and the two conditional properties, we can prove the functional properties of $\alpha$ to make the ratios micro-founded so as to explain the meaning of the signals contained in $\alpha(X)$. 

\textbf{Proposition 1:} \textit{Let} $\alpha$ \textit{follow the form as described in Eq. (7):}
\vspace{-0.5cm}
\begin{enumerate}
    \item \textit{If} $\theta(X) \geq \theta(Y)$ \textit{and} $\rho(X) < \rho(Y)\; \forall\; X,\;Y \in \mathcal{X}$ \textit{such that} $X \geq Y$, \textit{then} $\alpha$ \textit{satisfies the monotonicity property} (M).
    \item \textit{If} $\theta$ \textit{is concave and} $\rho$ \textit{convex, then} $\alpha$ \textit{satisfies the quasi-concavity property} (Q).
    \item $\rho({\lambda}X) = {\lambda}\rho(X)$ \textit{and} $\theta({\lambda}X) = {\lambda}\theta(X) \forall X \in \mathcal{X}\;and \;\lambda\in\;\mathbb{R}_{+} \backslash
    \{0\}\;such\;that\;{\lambda}X \in \mathcal{X}$, \textit{then} $\alpha$ \textit{satisfies the scale-invariance property} (S).
    \item If $\theta$ and $\rho$ \textit{satisfy the distribution-based property} (D), \textit{then so does} $\alpha$. 
\end{enumerate}

\textbf{Proof} is straightforward and mentioned in \citet{cheridito-kromer}. 

One of the key issues when measuring the R/R ratios over the revised VIX is that while computing performance ratios over a convex set of random variables generates independent and identically distributed (i.i.d.) variables, the financial return itself is not i.i.d., so this hinders the process of identifying uncertainty shocks as i.i.d. To mitigate this, we adopt the method of fitting a fractional time series model to take into account the long memory of the mean and volatility exhibited in the time series data. \citet{BBM} introduce the FIGARCH (Fractionally Integrated GARCH) model, demonstrating that traditional GARCH models are inadequate for capturing long memory in volatility. This finding highlights the need for fractional integration in volatility modeling to better reflect persistent effects in financial time series. This justifies the need for fractional integration in volatility modeling. Similarly, \citet{hyung-franses} shows that long memory in both the mean and variance processes is better modeled and captured using Autoregressive Fractionally Integrated Moving Average-Fractionally Integrated GARCH (ARFIMA-FIGARCH) models. Hence, the goal of the present paper is to emphasize the use of fractional time series models to capture the long memory that is explained by the multiple subordinated NIG L\'{e}vy process

Figure 8 gives an illustration of the difference between the innovations of the ARFIMA(1, $d(m)$, 1)-FIGARCH(1, $d(v)$, 1) and the Autoregressive Moving Average-GARCH of lag 1 and order 1 (ARMA(1,1)-GARCH(1,1)) fitted over the values of the newly constructed normalized VVIX. The fractional time series model is better at capturing the persistent effects created by the shocks implied by the newly constructed volatility index.

\begin{figure}[H]
    
    \includegraphics[width=1\linewidth]{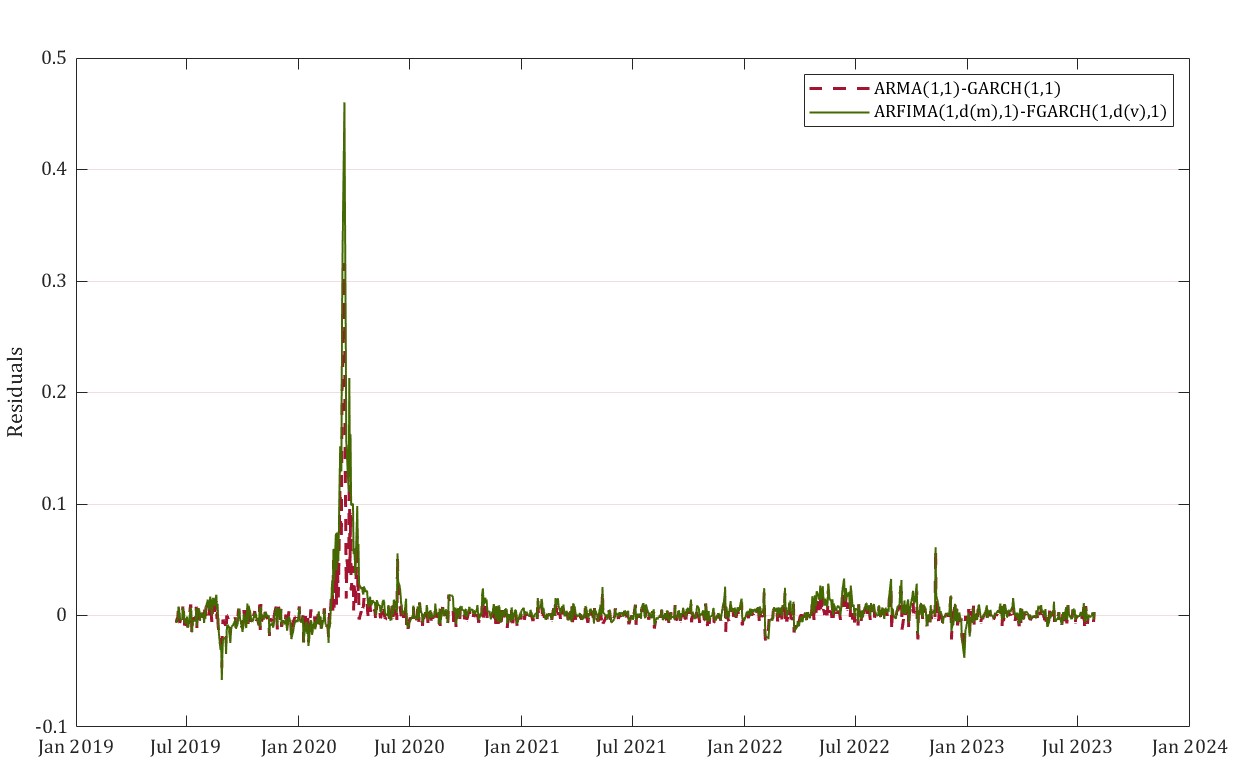}
    \caption{Residuals of the fitted time series models}
    \label{9}
\end{figure}

The long memory in the mean, captured using the ARFIMA model, refers to the persistence of past values of a time series influencing future values over long periods. In financial time series, long memory in the mean implies that past values of the series have a significant, slowly decaying influence on future values. Therefore, innovations to the time series do not fade away quickly, but explain the influence on the mean for a long time. ARFIMA models allow a slower, hyperbolic decay, characterizing a long memory. In addition, if a time series has long memory in its volatility, meaning the persistence of past volatility (variance) over time, large shifts in volatility appear to cluster and stay accentuated\ for long periods before decaying to normal levels. Long memory of volatility is present in financial markets (financial time series), where periods of high volatility (e.g., during a financial crisis) tend to last for extended periods and generate persistent shocks. 

Therefore, to capture the long memory of the mean and volatility exhibited by the time series of the normalized VVIX constructed in this paper, we apply the ARFIMA(1, $d(m)$, 1)-FIGARCH(1, $d(v)$, 1), where $d(m)$ is the term describing the long memory of the mean and $d(v)$ is the term describing the long memory of the volatility. The time series follows the process:

\begin{equation}
    ARFIMA(1,\;d(m),\;1) \rightsquigarrow {\phi}(L)(1 - L)^{d(m)}{z_{t}} = {\theta}(L){\varepsilon_{t}}
\end{equation}

In Eq. (14), $L$ is the lag operator, $d(m)$ is the fractional differencing parameter, reported to be 0.268, $\phi(L)$ is the autoregressive polynomial, while $\theta(L)$ is the MA polynomial. $z_{t}$ contains $n\times 1$ values of the normalized VVIX, and $\varepsilon_{t}$ is the $n\times 1$ vector of white noise error term.

\begin{equation}
    FIGARCH(1,\;d(v),\;1) \rightsquigarrow {\phi}(L)(1 - L)^{d(v)}{\varepsilon_{t}^{2}} = \omega + [1 - \beta(L)]\nu_{t}
\end{equation}

In Eq. (15), $\phi(L)$ is the autoregressive polynomial, and $d(v)$ is the fractional differencing parameter for volatility, reported to be 0.01. $\varepsilon_{t}^{2}$ is the square of the white noise error term to capture the conditional variance generating persistent volatility. $\omega$ is the constant term and $\beta(L)$ is the lag polynomial. Lastly, $\nu_{t}$ is the $n \times 1$ vector of normal innovations. To allow for a long memory in the fractional time series, we set the condition $d>0$. In cases where $d = 0$, the model is a standard ARMA(1,1)-GARCH(1,1) process.

Furthermore, it is essential to determine whether there is a predictable signal in the noise, as defined by the performance ratios, in the innovations that can make the markets inefficient given that using this measure of the revised VIX, agents will be able to forecast volatility price. To detect the predictable signal in the noise, we simulate $S = 10,000$ scenarios of the normalized VVIX (with NDIG distribution) over the ARFIMA(1, $d(m)$, 1)-FIGARCH(1, $d(v)$, 1) process as defined by Eqs. (14) and (15). We compute the Rachev ratio and the Stable Tail Adjusted Return ratio\footnote{The functional forms of these performance ratios are described in Section 7.} over $S$ scenarios to extract predictable signals from the volatility noise. Figure (12) shows the performance ratios computed over the simulated scenarios. 

\begin{figure}
 \begin{minipage}[b]{0.49\textwidth}
        \centering
        \includegraphics[width=\textwidth]{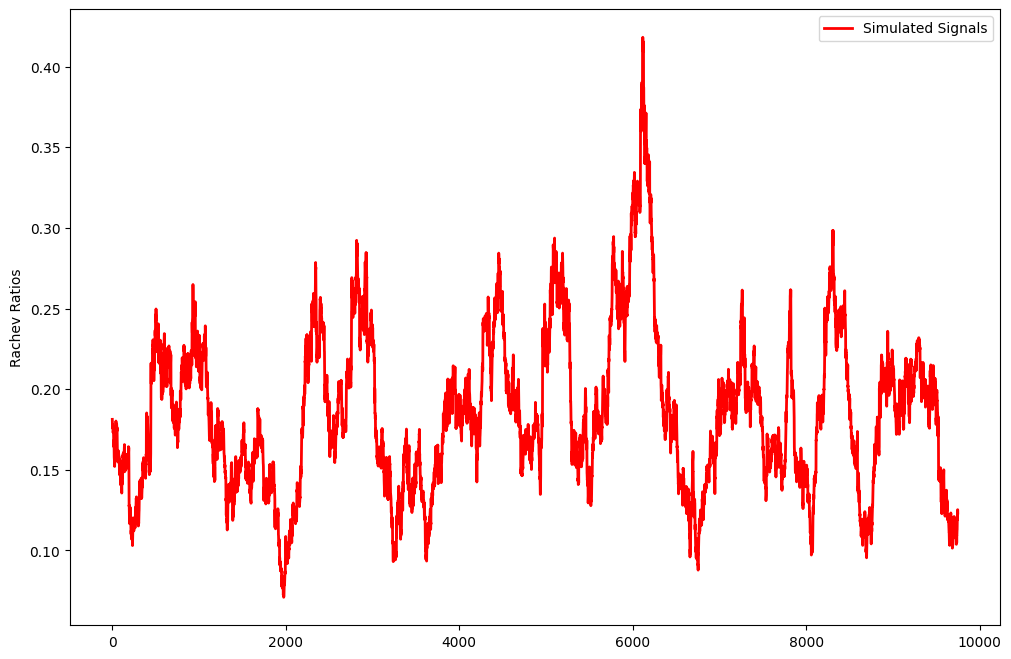}
    \end{minipage}
    \hfill
    \begin{minipage}[b]{0.49\textwidth}
        \centering
        \includegraphics[width=\textwidth]{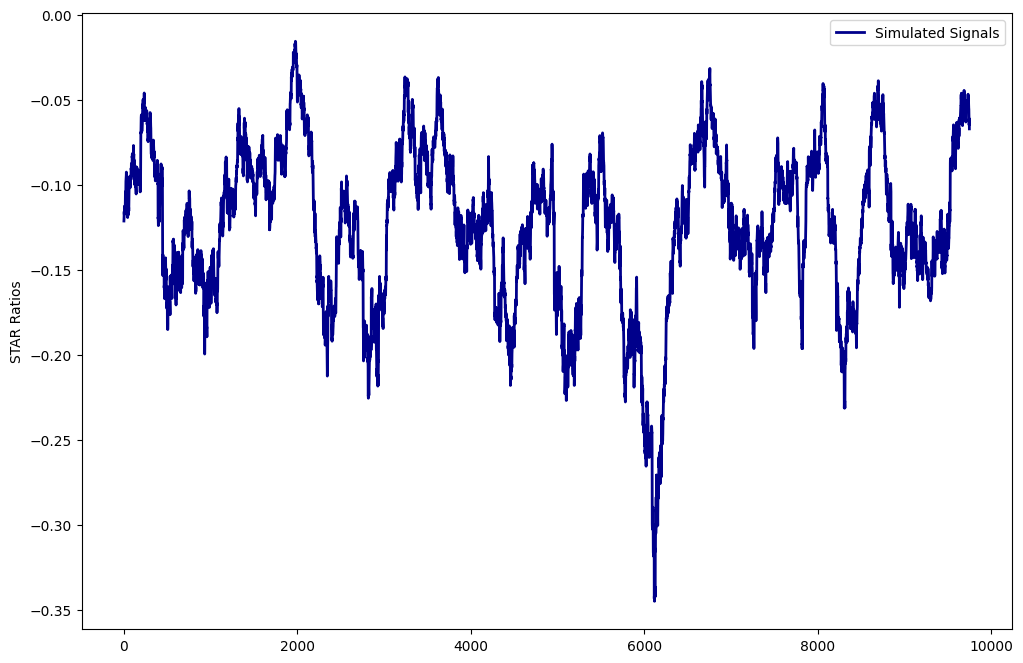}
    \end{minipage}
    \caption{\centering{Signal/Noise ratios detected using performance ratios over $S$ scenarios}}
    \label{10}
\end{figure}

From the simulated signals using the fractional time series process, it is evident that there is significant randomness in the volatility noise with volatility clustering and generates no predictable power. Given that there is no identifiable pattern in the volatility noise that can enable an agent to forecast volatility price, we can conclude that the revised measure of VIX generates randomness in volatility noise and satisfies the Efficient Market Hypothesis.

\section{Identification Strategy}
\setlength{\parskip}{10pt}
\justifying

To identify i.i.d. shocks, using the normal innovations extracted by utilizing Eqs. (14) and (15), we compute R/R ratios over the residuals of the ARFIMA(1, $d(m)$, 1)-FIGARCH(1, $d(v)$, 1) process. For illustrative purposes, we compute two performance ratios\footnote{The phrases `R/R ratios' and `performance ratios' are used interchangeably.} namely, the Rachev ratio and STAR ratio over the normal innovations. The following are the functional forms of the two ratios.

\begin{enumerate}
    \item \textit{Rachev Ratio:} 
    
    \begin{equation}
        RR_{(\beta,\;\gamma)}(X)\;:=\frac{AVaR_{\beta}(-X)}{AVaR_{\gamma}(X)}
    \end{equation}

\noindent     where ${AVaR_{\beta,\;\gamma}(X)}:= \beta^-1\int_{0}^{\beta}[max(-F_{x}^{-1}(u),0)]^{\gamma}du$, where $AVaR$ is defined as the Average Value at Risk and $X$ is the measure of interest, in this case, normal innovations of the revised VIX. $\beta$ refers to the confidence interval of the value on the right tail, whereas $\gamma$ refers to the confidence interval of the value on the left tail. While the Rachev ratio satisfies the properties (M), (S), and (D), it violates (Q) due to a non-concave numerator. 
\vspace{0.5cm}

    \item \textit{Stable Tail Adjusted Return Ratio (STAR Ratio):}

    \begin{equation}
        STARR_{\gamma}(X) := \frac{\mathbb{E}[X]^+}{AVaR_{\gamma}(X)^+} 
    \end{equation}

    \noindent where $AVaR_{\gamma}(X) := \gamma^{-1}\int_{0}^{\gamma}VaR_{u}(X)du$ is the Average-Value-at-Risk at the level $\gamma \in (0,1].$ STARR satisfies all four axioms namely (M), (Q), (S), and (D), therefore, is axiomatically robust.
\end{enumerate}

Eqs. (16) and (17) will be used as the benchmark performance ratios for computing the uncertainty shocks (signals from the volatility noise) that follow the ARFIMA(1, $d(m)$, 1)-FIGARCH(1, $d(v)$, 1) process. Therefore, Figure (13) shows the performance ratios computed over the normal innovations of the fitted fractional time series model as the new i.i.d series of uncertainty shocks. 

\begin{figure}[H]
 \begin{minipage}[b]{0.49\textwidth}
        \centering
        \includegraphics[width=\textwidth]{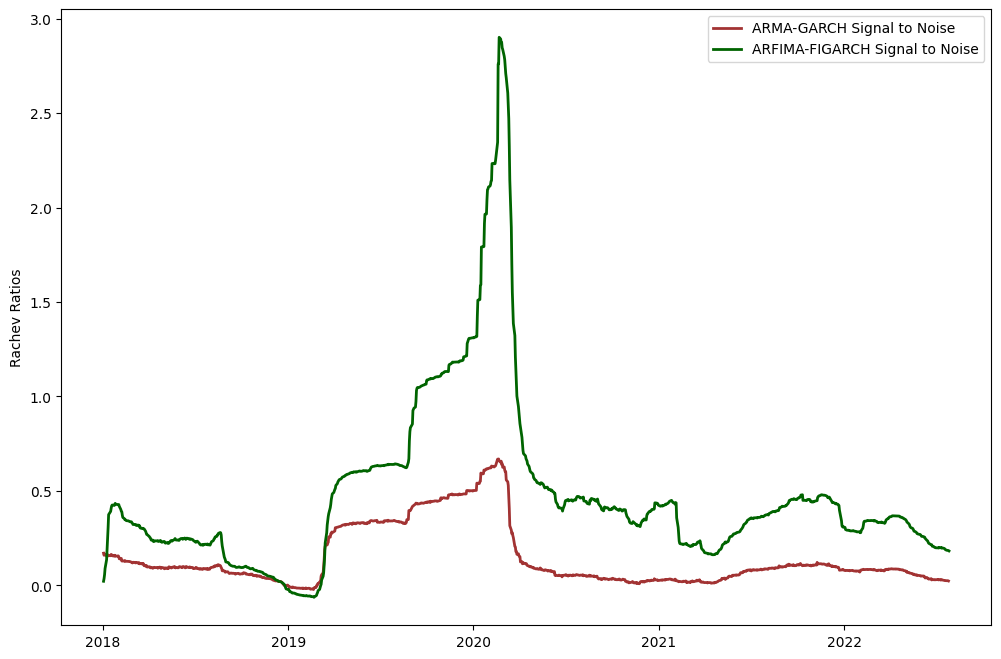}
    \end{minipage}
    \hfill
    \begin{minipage}[b]{0.49\textwidth}
        \centering
        \includegraphics[width=\textwidth]{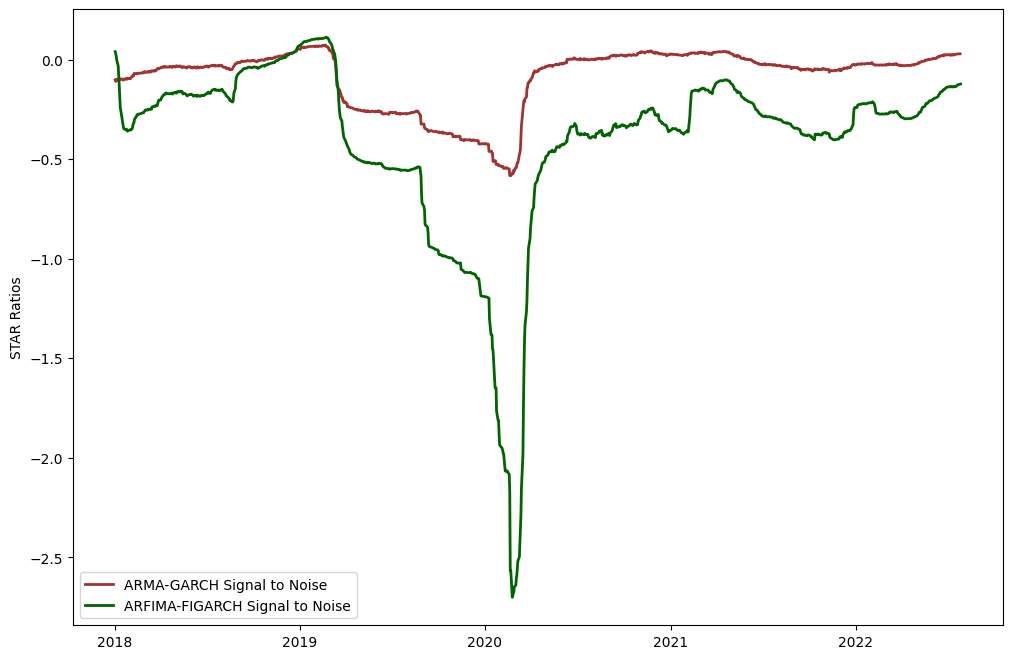}
    \end{minipage}
    \caption{\centering{Novel Uncertainty Shocks. (i) Rachev Ratios (ii) STAR Ratios}}
    \label{11}
\end{figure}

Our uncertainty shocks \textit{(ARFIMA-FIGARCH Signal to Noise)} are not serially correlated, as confirmed by Figure (12). Hence, the assumption that macroeconomic shocks are uncorrelated over time can be defended by arguing that there are no identifiable signals about volatility in the noise (demonstrated in the scenario simulation) and do not depend on any past shocks. Since uncertainty is considered an unexpected and independent event, we can say that these uncertainty shocks are serially uncorrelated, and we account for volatility clustering using the long memory of the fractional time series and the integration of the double subordinated NIG L\'{e}vy process fit to the option prices of the S\&P 500. 

Intuitively, it is reasonable to argue that the large steps or jumps in the uncertainty shocks observed are detectable signals, particularly when using R/R ratios to detect the signal-to-noise ratio from the time series. Moreover, performance ratios might capture underlying financial market dynamics differently from standard methods, making these shocks more prominent or close to the true magnitude of financial market volatility as a proxy for uncertainty. These jumps or steps could reflect the underlying market conditions more robustly, particularly while measuring the impact of uncertainty or volatility on the behavior of the financial market. 

Performance ratios 1 and 2 focus on the aspect of the signal-to-noise ratio, which allows us to highlight fluctuations in the financial market that traditional local volatility models might miss. The robustness comes from the ability of our model to capture these dynamics with more sensitivity to changes that directly affect market participants' risk--reward trade-offs. In addition, this identification is compelling because the innovations (using ARFIMA-FIGARCH) account for persistent and long-memory effects, which are often observed in financial market data. Therefore, these larger jumps could be indicative of shifts in the market's perception of risk and volatility in response to significant events or structural changes in the economy.

Furthermore, these uncertainty shocks are equipped to explain several key events, alluded to in section 1, in the sample period over which they have been computed. In Figures (13 (i) and (ii)), we can see the shock occurring in March 2020 which explains the major plummet of the S\&P500 index due to the news about COVID-19 forcing lock-downs across the United States along with major developed countries. This period saw the financial market crashing significantly, leading to heightened uncertainty about economic activities and triggering the risk-aversion behavior of the agents. Moreover, it is intuitively safe to assume that these uncertainty shocks explain the effect of risk-averse agents, in complete markets, liquidating their holdings in risky assets and transferring wealth into risk-free assets to enable precautionary savings consistent with the consumption risk-sharing hypothesis.

Our strategy for identifying uncertainty shocks, particularly during early 2022, leverages significant movements in the S\&P 500 index driven by multiple interrelated factors. This period marks the second major event in recent financial history where elevated volatility and uncertainty shocks are observable, as captured by our model. The key drivers of these shocks include macroeconomic conditions, geopolitical risks, and sector-specific factors, which align with both theoretical frameworks and empirical observations.

First, rising inflationary pressures in early 2022, exacerbated by supply chain disruptions and a post-pandemic rebound in global demand, led the Federal Reserve to signal an aggressive tightening of monetary policy. This response, aimed at fulfilling the Fed's dual mandate of stabilizing prices and maximizing employment, introduced heightened uncertainty into financial markets. The market response was reflected in sharp declines in the S\&P 500, particularly as the trajectory of interest rates became uncertain. This aligns with the established literature on monetary policy uncertainty, where market participants react strongly to the ambiguity surrounding future rate hikes and their potential impact on risky assets (see, e.g., \citet{bloomfu} and \citet{jurado2015}). The forward guidance provided by the Federal Reserve increased volatility as markets began to price-in the risks of a more restrictive policy environment, leading to higher risk premia and a greater sensitivity to macroeconomic news.

Second, the geopolitical shock arising from Russia's invasion of the Ukraine in February 2022 serves as a clear catalyst for heightened uncertainty. Geopolitical events, such as armed conflicts, are known to produce large, exogenous shocks to the economy, often characterized as rare disaster events within the framework of tail risk (see \citet{barro} and \citet{routledge}). These events significantly impact asset prices due to the sudden and unpredictable nature of the disruptions they cause to global trade, energy markets, and investor sentiment. Our analysis demonstrates that these geopolitical shocks are captured by the fat-tailed behavior of the S\&P 500 distribution, which reflects a shift in the risk-neutral density, consistent with the rare disaster hypothesis. Such tail risks are not adequately captured by traditional measures of market volatility alone but are crucial for understanding the full scope of uncertainty shocks in periods of geopolitical crisis.

Third, the sharp correction in high-growth technology stocks during early 2022 provides another dimension to the uncertainty shocks identified in our model. Many of these firms had experienced meteoric rises during the pandemic due to favorable liquidity conditions and investor expectations of continued high growth. However, as inflation and interest rates increased, the present value of these firms' future cash flows was discounted more heavily, leading to sharp declines in their valuations. This sectoral shock was compounded by low earnings reports from several Fortune 500 technology companies, which introduced additional uncertainty at the firm level. Firm-level uncertainty, particularly in sectors like technology, is often driven by earnings volatility and future profitability concerns, as outlined by \citet{bloom2007}. The declines in these stocks reflected broader concerns about the sustainability of growth in the face of rising costs and tightening monetary conditions, further amplifying the aggregate uncertainty in financial markets.

Our identification strategy highlights the multifaceted nature of the uncertainty shocks in financial markets by identifying effects arising out of macroeconomic shocks, geopolitical risk, and sectoral disruptions. The use of the S\&P 500 as a proxy for these shocks is well-supported by its role as a barometer of overall market sentiment and risk appetite. Additionally, by incorporating insights from the rare disaster literature and firm-level uncertainty frameworks, our analysis captures both systemic and idiosyncratic factors contributing to market-wide uncertainty during this period.

\section{Conclusion}

\setlength{\parskip}{10pt}
\justifying

This paper presents a novel approach to identifying uncertainty shocks in financial markets, focusing on the heavy-tailed, non-Gaussian nature of asset returns. By fitting a double-subordinated Normal Inverse Gaussian (NIG) L\'{e}vy process to S\&P 500 option prices, we constructed a more robust measure of volatility, the Volatility of VIX (VVIX). This revised VIX captures large market movements incorporating features like skewness and heavy tails commonly observed in financial data.

Our methodology extends the traditional framework for measuring uncertainty by introducing a general family of R/R ratios. These ratios, computed on the fitted fractional time series of the revised VIX, offer a more nuanced understanding of the relationship between risk and return in the presence of extreme tail risks. This approach not only improves the identification of uncertainty shocks but also enhances our ability to model macroeconomic volatility and its impact on financial markets.

In summary, our findings suggest that standard second-moment measures of volatility and using local volatility models to compute volatility surfaces of the VIX are insufficient for capturing the full extent of volatility (implying uncertainty) in financial markets, especially during periods highlighting extreme tail risk. The proposed VVIX, derived from a double-subordinated NIG L\'{e}vy process, provides a more accurate representation of market volatility and its associated risks, making it a valuable tool for both researchers and practitioners in economics and finance to explore new avenues concerning the impact of uncertainty shocks. Lastly, future research may explore the application of this method to other classes of assets and its implications for portfolio management, risk mitigation strategies, and more importantly, to better understand the true effect of uncertainty on macroeconomic indicators.


\pagebreak
\bibliography{references_clean}

\pagebreak
\section*{Appendix}
\setlength{\parskip}{10pt}

\begin{figure}[H]
    \centering
    \includegraphics[width=1\linewidth]{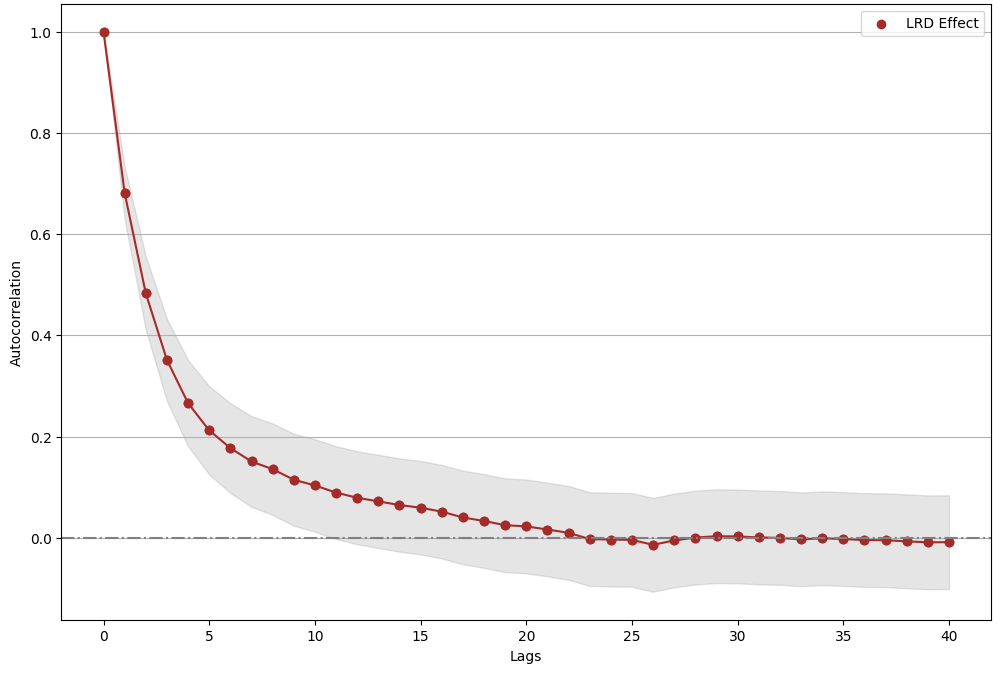}
    \caption{Long range dependences: Decay of the autocorrelation}
    \label{12}
\end{figure}

Figure (14) displays the autocorrelation function of the residuals from an ARFIMA-FIGARCH model fitted to the revised VIX, revealing long-range dependencies characteristics. The slow decay of the autocorrelation indicates that volatility is highly persistent, as the residuals show significant autocorrelation even beyond the initial lags. The decay begins sharply but gradually flattens, with the autocorrelation remaining positive up to about 22 lags before approaching zero. This suggests that the ARFIMA-FIGARCH model captures the persistent volatility patterns inherent in the VIX, effectively modeling the heavy-tailed nature and memory effects associated with market uncertainty. The lag at which the autocorrelation converges close to zero implies that the model successfully accounts for the long memory in the volatility, which is crucial for understanding the dynamics of financial market stress.

Figure (15) displays the STAR ratio computed over the normal innovations of the current VIX (using ARMA(1,1)-GARCH(1,1)) and the normal innovations of the revised VIX (VVIX) (using ARFIMA(1, $d(m)$, 1)-FIGARCH(1, $d(v)$, 1)). The new measure of uncertainty shocks is capturing more pronounced signals about financial market volatility than the measure extracted from the current VIX. Given that fitting an ARFIMA(1, $d(m)$, 1)-FIGARCH(1, $d(V)$, 1) on the current VIX has not been explored in the literature, we revert to the standard ARMA(1,1)-GARCH(1,1) when $d=0$.

\begin{figure}[H]
    \centering
    \includegraphics[width=1\linewidth]{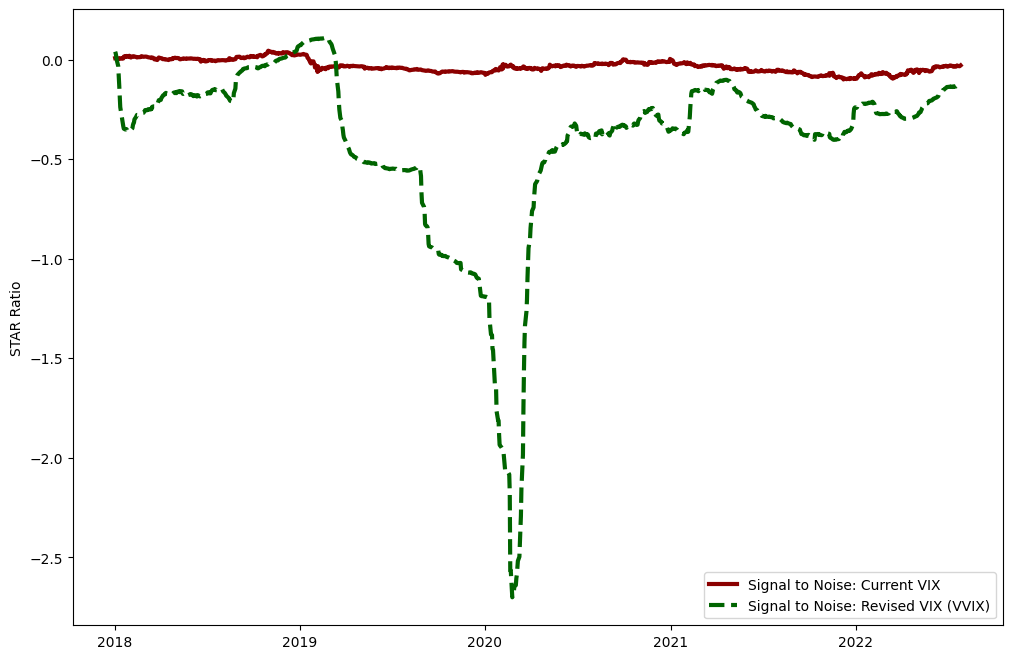}
    \caption{Signal to Noise: Current VIX v/s Revised VIX (VVIX)}
    \label{13}
\end{figure}

\end{document}